\documentclass[twocolumn, manuscript]{revtex4}
\usepackage{graphicx}
\begin{document}

\title{Natural line shape}
\author{P.V. Elyutin}
\email{pve@shg.phys.msu.su} \affiliation {Department of Physics,
Moscow State University, Moscow 119991, Russia}
\date{\today}

\begin{abstract}
The observable line shape of the spontaneous emission depends on
the procedure of atom's excitation.  The spectrum of radiation
emitted by a two-level atom excited from the ground state by a pi
pulse of the resonant pump field is calculated for the case when
the Rabi frequency is much larger than the relaxation rate. It is
shown that the central part of the spectral distribution has a
standard Lorentzian form, whereas for detunings from the resonance
that are larger than the Rabi frequency the spectral density falls
off faster.  The shape of the wings of the spectral line is
sensitive to the form of the pi pulse.  The implications for the
quantum Zeno effect theory and for the estimates of the duration
of quantum jumps are discussed.

\vspace{10mm} PACS numbers: {42.50.Ct, 42.50.Lc, 32.70.Jz}
\end{abstract}
\maketitle

We shall treat the problem of the natural line shape -  the shape
of the spectral line of the spontaneous emission - the radiation
emission by a secluded atom accompanying the transition from the
excited state (to some lower one) that originates from the
interaction of the atom with the quantized electromagnetic field.
In what follows we use the model of the two-level atom with the
excited state $\left| 2 \right\rangle $ and the ground state
$\left| 1 \right\rangle $ with the energies $E_2 $ and $E_1 $
respectively, which are connected by the electrical dipole
transition. The theory originally developed by Weisskopf and
Wigner \cite{WW30} starts with the assumption of the exponential
decay of the amplitude   of the initial state,
\begin{equation}\label{1}
B\left( t \right) = {\mathop{\rm e}\nolimits} ^{ - \gamma t}.
\end{equation}
It yields the Lorentzian line shape,
\begin{equation}\label{2}
S\left( \omega  \right) = \frac{1}{\pi } \cdot \frac{\gamma
}{{\left( {\omega _0  - \omega } \right)^2  + \gamma ^2 }}
\end{equation}
where $\omega _0  = {{\left( {E_2  - E_1 } \right)}
\mathord{\left/ {\vphantom {{\left( {E_2  - E_1 } \right)} \hbar
}} \right.  \kern-\nulldelimiterspace} \hbar }$ is the transition
frequency ($\hbar $ is the Planck's constant).  The natural
linewidth is  $\gamma  = {\Gamma  \mathord{\left/
 {\vphantom {\Gamma  2}} \right.  \kern-\nulldelimiterspace} 2}$,
where the transition rate $\Gamma $ is given by the equality
\begin{equation}\label{3}
\Gamma  = \frac{{2\pi }}{\hbar }\left| {V_{nk} } \right|^2 \rho
\left( {E_k } \right),
\end{equation}
that was originally derived by Dirac \cite{D27} and now is
universally known as the Fermi golden rule.  In Eq. (3) $n $ and
$k $ denote the initial and final states of the system "atom +
field", $V_{nk} $ are the matrix elements of the interaction of
atom with the quantized field, and $\rho (E_k) $  is the energy
density of the final states.  The summation over the quantum
numbers others than energy is carried out.

The expression (3) includes the values of the matrix elements and
of the density of states on the energy surface $E = E_n  = E_k $.
Since the relative variations of $V_{nk} $ and $\rho (E_k) $
within the main peak of the spectral line (2) in the optical range
have values about ${\gamma  \mathord{\left/{\vphantom {\gamma
{\omega _0 }}} \right. \kern-\nulldelimiterspace} {\omega _0
}}\sim 10^{-8} $, the Lorentzian line shape promises to be very
accurate.  To my knowledge, the deviations from (2) were never
established experimentally.

However, from the theoretical point of view the Lorentzian form of
spectrum is irritating.  On one side, this expression can not be
universally valid, since the negative frequencies of photons are
physically meaningless.  On the other side, Eq. (2) gives for the
mean frequency of the radiation $\overline {\omega} $ the
expression that diverges - literally logarithmically, but even
faster, if the energy density growth $\rho  \propto \omega ^2 $ is
taken into account \cite{MCA89}.

The line shape Eq. (2) can be interpreted as the form of the
energy distribution $W\left( E \right) = \hbar ^{ - 1} S\left(
{\hbar \omega } \right)$  for the quasistationary initial state of
the system $\left| {\Psi \left( 0 \right)} \right\rangle $.  The
law of decay of the initial state is given by the Fourier
transform of the energy distribution,
\begin{eqnarray}\label{4}
\Phi \left( t \right) = \left| {\left\langle {{\Psi \left( 0
\right)}} \mathrel{\left | {\vphantom {{\Psi \left( 0 \right)}
{\Psi \left( t \right)}}} \right. \kern-\nulldelimiterspace}
{{\Psi \left( t \right)}} \right\rangle } \right|^2 \quad \\
\nonumber\qquad  = \left| {\,\int {W\left( E \right)\exp \left( {
- i\frac{{Et}}{\hbar }} \right)dE} \,} \right|^2.
\end{eqnarray}
The Lorentzian form of the energy distribution leads to the
exactly exponential decay law $\Phi \left( t \right) = e^{ -
\Gamma t} $, that makes the Weisskopf and Wigner theory
self-consistent. Thence the problem of the natural line shape can
be reformulated as the problem of the law of the decay of the
initial state: deviations from Eq. (2) will lead to the
nonexponentiality of the decay - and vice versa.  The additional
incentive to study the detailed form of the initial stage of the
decay law came from the concept of the quantum Zeno effect
\cite{MS77, CSM77}.  If for small times the decay law is
quadratic, $\Phi \left( t \right) \approx 1 - \left( {{t
\mathord{\left/ {\vphantom {t {\tau _Z }}} \right.
\kern-\nulldelimiterspace} {\tau _Z }}} \right)^2 $ (the parameter
$\tau _Z $ is known as the Zeno time), then frequent measurements
of the energy of the system will prevent the decay of the initial
excited state. This property permitted Schulman \cite{Sch97} to
introduce the estimate for the duration of the quantum jump
between the atomic states as the crossover time from the quadratic
decay to the exponential (Fermi) decay:
\begin{equation}\label{5}
\tau _J  = \Gamma \tau _Z^2.
\end{equation}
The logic behind this definition is lucid: if the measurement can
influence the kinetics of the quantum jump, then it has not been
completed yet.

Several authors have attempted to calculate the Zeno time $\tau _Z
$ for the spontaneous emission of radiation \cite{H81, Sch97,
FP98}. They have used the two-level model of the atom; in this
case the account of the momentum of the emitted photon suppresses
the matrix elements in the high frequency domain and effectively
truncates the Lorentzian line shape at the frequencies around
$\omega _ + = \alpha ^{ - 1} \omega _a $, where $\alpha  = {{e^2 }
\mathord{\left/ {\vphantom {{e^2 } {\hbar c}}} \right.
\kern-\nulldelimiterspace} {\hbar c}}$ is the fine structure
constant ($e $ is the electric charge of the electron and $c $ is
the speed of light), and the atomic frequency unit $\omega _a  =
me^4 \hbar ^{ - 3} $ ($m $ is the electron's mass).  However, the
results of these authors contradict their assumptions: if the
hydrogen atom in the initial state $2p $  can indeed emit a photon
with the energy about $E_ +   = \hbar \omega _ +   =
3.7\,{\rm{keV}}$, then there is no reason to limit the channels of
decay by the transition only to the state $1s $.  With the account
of all possible transitions the dispersion $\Delta \omega ^2 $ of
the energy distribution of the initial state diverges; that
prohibits the existence of the quadratic stage of the decay and
makes $\tau _Z = 0$.

We note that the influence of the truncation of the Lorentzian
shape (2) in the domain $\omega <0 $ on the decay law has been
studied in Ref. \cite{SG91}.  This truncation diminishes the
initial decay rate by half, but doesn't lead to the quadratic
decay.

The physically unsatisfactory divergences of $\overline \omega $
and $\Delta \omega ^2 $ are rooted in the unphysical initial
conditions.  It is a commonplace of the theory of quasistationary
states that their properties depend on the procedure of their
preparation \cite{Kh90}.  Therefore this procedure must be taken
into account explicitly.  The importance of this approach in the
problem of the natural line shape was noted long time ago
\cite{L52}.

Let's assume that the atom initially was in the ground state
$\left| 1 \right\rangle $ and then was excited by a pulse of the
resonant pump field (of the frequency $\omega _0 $) that has the
properties of the pi pulse \cite{AE75}, that conveys the two-level
system (in the absence of relaxation) from one state into the
other. We take the Hamiltonian of the system in the form
\begin{equation}\label{6}
\hat H = \hat H_a  + \hat V_q  + \hat V_c  + \hat H_f ,
\end{equation}
where $\hat H_a $ and $\hat H_f $ are the Hamiltonians of the
two-level atom and of the quantized field respectively, and $\hat
V_q $ and $\hat V_c $ take account of the atom's interaction with
the quantized radiation field and with the classical pump field
correspondingly. In the dipole approximation we have
\begin{equation}\label{7}
\hat V_q  = \sum\limits_\lambda  {\hat v_\lambda  }
,\,\,\,\,\,\,\hat v_\lambda   =  - \frac{e}{m}\sqrt {\frac{{2\pi
\hbar }}{{L^3 \omega _\lambda  }}} {\mathbf{\hat pe}}_\lambda
\left[ {\hat a_\lambda ^ +   + \hat a_\lambda } \right]
\end{equation}
where the index $\lambda $ numerates the modes of the quantized
field.  Here $L^3 $ is the quantization volume, ${\mathbf{\hat
p}}$ is the operator of the momentum of the atomic electron,
${\omega _\lambda  }$ and ${\mathbf{e}}_\lambda  $ are the
frequency of the photon and the polarization vector of the mode
$\lambda $ respectively, and ${\hat a_\lambda ^ +  }$ and ${\hat
a_\lambda }$ are the creation and annihilation operators for this
mode.  The operator $\hat V_c $ is
\begin{equation}\label{8}
\hat V_c  =  \frac{e}{{m\omega _0 }}{\mathbf{\hat pE}}\left( t
\right)\cos \omega _0 t,
\end{equation}
where ${\mathbf{E}}\left( t \right)$ is the envelope of the
electric field of the pulse.

The state vector of the system can be expanded as
\begin{eqnarray}\label{9}
\left| {\Psi \left( t \right)} \right\rangle  = A\left|
{1,{\rm{V}}} \right\rangle e^{ - i\omega _1 t}  + B\left|
{2,{\rm{V}}} \right\rangle e^{ - i\omega _2 t}  \\
\nonumber +\sum\limits_\lambda  {C_\lambda  \left| {1,\lambda }
\right\rangle e^{ - i\left( {\omega _1  + \omega _\lambda  }
\right)t} },\quad \quad \,\,\,\,\,
\end{eqnarray}
where $\left| {j,{\rm{V}}} \right\rangle $ denote states of the
system with the atom in the state $\left| j \right\rangle $ and
the field in the vacuum state; in the state $\left| {1,\lambda }
\right\rangle $ the atom is in the ground state $\left| 1
\right\rangle $, one photon is present in the mode $\lambda $, and
there are no photons in other modes; $\omega_j = E_j/\hbar $.

The evolution of the system is governed by the system of equations
\begin{equation}\label{10}
i\frac{{dA}}{{dt}} = B\Omega \left( t \right)\cos \omega_0 t
e^{-i\omega _0 t},
\end{equation}
\begin{equation} \label{11}
i\frac{{dB}}{{dt}} = A\Omega \left( t \right)\cos \omega_0 t
e^{i\omega _0 t}+\sum\limits_\lambda {u_\lambda C_\lambda
e^{i\Delta _\lambda t} } ,
\end{equation}
\begin{equation} \label{12}
\\i\frac{{dC_\lambda }}{{dt}} =
u_\lambda Be^{ - i\Delta _\lambda t} .
\end{equation}
Here $\Omega \left( t \right)$ is the (time-dependent) Rabi
frequency, $\Omega \left( t \right) = {{e{\bf{\hat p}}_{12}
{\bf{E}}\left( t \right)} \mathord{\left/ {\vphantom
{{e{\mathbf{\hat p}}_{12} {\mathbf{E}}\left( t \right)} {m\omega
_0 \hbar }}} \right. \kern-\nulldelimiterspace} {m\omega _0 \hbar
}}$, where ${\mathbf{\hat p}}_{12} $ is the matrix element of the
momentum between the states $\left| 1 \right\rangle $ and $\left|
2 \right\rangle $,
\begin{equation}\label{13}
u_\lambda   =  - \frac{e}{m}\sqrt {\frac{{2\pi }}{{L^3 \hbar
\omega _\lambda  }}} {\bf{p}}_{12} {\bf{e}}_\lambda,
\end{equation}
and $\Delta _\lambda   = \omega _0  - \omega _\lambda  $ is the
frequency detuning between the atomic transition and the emitted
photon.

Let's consider the pi pulse with the rectangular envelope,
\begin{eqnarray}\label{14}
\Omega \left( t \right) = \Omega \quad {\rm{ }}\left( { -
\frac{\pi
}{\Omega } < t < 0} \right),\\
\nonumber\,\,\,\Omega \left( t \right) =
0\,\,\,\,\qquad{\rm{otherwise}},
\end{eqnarray}
and assume that the Rabi frequency is much larger than the
relaxation rate, $\Omega \gg \gamma $. Then throughout the
duration of the pulse we can neglect the spontaneous radiation,
and take into account only the pump field.  Thus from Eqs. (10)
and (11) with the initial conditions $A\left( { - {\pi
\mathord{\left/ {\vphantom {\pi \Omega }} \right.
\kern-\nulldelimiterspace} \Omega }} \right) = 1$, $B\left( { -
{\pi  \mathord{\left/ {\vphantom {\pi \Omega }} \right.
\kern-\nulldelimiterspace} \Omega }} \right) = 0$ in the rotating
wave approximation we obtain
\begin{equation}\label{15}
B\left( t \right) =  - i\sin \frac{\Omega
}{2}t\,\,\,\,\,\,\,\,\,\,\left( { - \frac{\pi }{\Omega } < t < 0}
\right).
\end{equation}
To describe the second stage, that of the free emission, we can
use the exponential decay of Eq. (1):
\begin{equation}\label{16}
B\left( t \right) =  - ie^{ - \gamma t}
\,\,\,\,\,\,\,\,\,\,\,\,\,\,\,\,\,\left( {t > 0} \right).
\end{equation}
By substitution of Eqs. (15) and (16) in Eq. (12) and integration
we obtain for the limiting values of amplitudes $C_\lambda  $:
\begin{equation}\label{17}
C_\lambda  \left( \infty  \right) = -u_\lambda  F\left( {\omega
_\lambda  } \right),
\end{equation}
where the spectral amplitude is given by the expression
\begin{equation}\label{18}
F\left( \omega  \right) = \frac{{2\Omega \exp \left( {i\frac{{\pi
\Delta }}{\Omega }} \right) - 4i\Delta }}{{\Omega ^2  - 4\Delta ^2
}} + \frac{1}{{i\Delta  + \gamma }}.
\end{equation}
The spectral distribution of photons is  $S\left( \omega  \right)
= N\left| {F\left( \omega  \right)} \right|^2 $ , where $N $ is
the normalization constant; in our case $N \approx {\gamma
\mathord{\left/ {\vphantom {\gamma  \pi }} \right.
\kern-\nulldelimiterspace} \pi }$.  The explicit expression for
$S\left( \omega  \right)$ is too cumbersome; it is more convenient
to work with the formula (18).

Firstly, it must be noted that the first term in the RHS of (18)
is regular at $\Delta  =  \pm {\Omega  \mathord{\left/ {\vphantom
{\Omega  2}} \right. \kern-\nulldelimiterspace} 2}$, since at
these points both the numerator and the denominator have simple
zeroes.  Secondly, for small frequency detunings $|\Delta|
\lesssim \Omega $ the second term in the RHS of (18) dominates,
and the line shape is given just by the standard Lorentzian form
(2). Thirdly, for large $|\Delta | \gg \Omega  \gg \gamma $, after
expanding both terms in negative powers of $\Delta $, we find that
two terms of the order $\Delta ^{ - 1} $ cancel each other. The
dominating contribution comes from the term
\begin{equation}\label{19}
F\left( \omega  \right) \approx  - \frac{\Omega }{2}\exp \left(
{i\frac{{\pi \Delta }}{\Omega }} \right)\Delta ^{ - 2} ,
\end{equation}
that defines the asymptotics of the spectral density
\begin{equation}\label{20}
S\left( \omega  \right) \approx \frac{\gamma }{{4\pi
}}\frac{{\Omega ^2 }}{{\Delta ^4 }}.
\end{equation}
It decreases rapidly enough to provide a finite average value of
the frequency of the emitted photons $\overline \omega  $ (that in
our approximation is indistinguishable from the transition
frequency $\omega _0 $) and the finite value of the frequency
dispersion $\Delta \omega ^2 \approx 0.39\Omega \Gamma $.

It must be noted that the behavior of the wings of the spectral
line depends on the form of the envelope of the pi pulse.  We have
calculated the spectral density $S (\omega ) $ also for the pi
pulse with the sine envelope:
\begin{eqnarray}\label{21}
\Omega \left( t \right) =  - \frac{\pi }{2}\Omega \,\sin
\Omega t{\rm{ }}\left( { - \frac{\pi }{\Omega } < t < 0} \right),\\
\nonumber \Omega \left( t \right) = 0\,\,\,\,
{\rm{otherwise}}{\rm{.}\qquad\,\,\,}
\end{eqnarray}
The results are compared in Fig. 1 with the Lorentzian line shape
and the spectral distribution for the rectangular envelope.  It
can be seen that for both types of pulses the crossovers between
Lorentzian and asymptotic forms occur at $\Delta \sim \Omega $.

\begin{figure}[!ht]
\includegraphics[width=1.0\columnwidth]{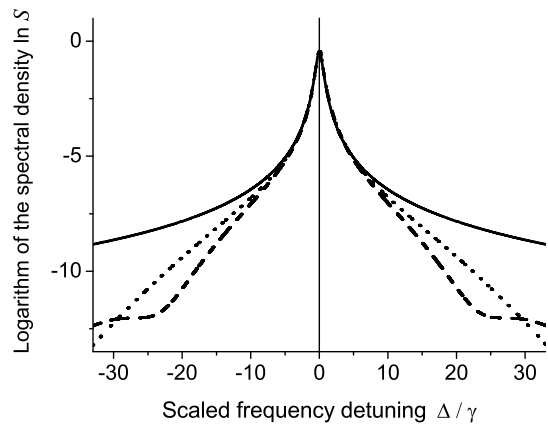}
\caption{\label{fig1}The dependence of the logarithm of the
spectral density $\ln \,S$ on a scaled frequency detuning $\Delta
/ \gamma $ for the Rabi frequency $\Omega  = 10\gamma $.  Solid
line - the Lorentzian form (2), dashed line corresponds to the
rectangular envelope of the pi pulse, Eq. (14), dotted line - the
same for the sine pulse envelope, Eq. (21).}
\end{figure}

The finite dispersion of the frequency defines the Zeno time of
the system \cite{Sch97}; thus $\tau _Z  = \left( {\Delta \omega ^2
} \right)^{{{ - 1} \mathord{\left/ {\vphantom {{ - 1} 2}} \right.
\kern-\nulldelimiterspace} 2}} \sim \left( {\Omega \Gamma }
\right)^{{{ - 1} \mathord{\left/ {\vphantom {{ - 1} 2}} \right.
\kern-\nulldelimiterspace} 2}} $.  Then from Eq. (5) we obtain a
somewhat trivial estimate for the duration of the quantum jump in
our case, $\tau _J \sim \Omega ^{ - 1} $.  It must be noted that
indefinite increase of the amplitude of the pi pulse will
eventually violate the applicability of the two-level model.  The
transitions to other states of the system are necessarily
important if the Rabi frequency is comparable to the transition
frequency, $\Omega \sim\omega _0 $.  Thus our model assures that
the duration of the quantum jump accompanying the spontaneous
emission from a given transition will always be larger than the
field period, $\tau _J \gtrsim \omega _0^{ - 1} $. This inequality
is almost universally accepted by the community of physicists on
the grounds of common sense and is reflected in the literature
\cite{IH+90}.

Our analysis is limited by the domain $\Omega  \gg \gamma $. It is
interesting to compare it with the opposite limiting case. For
$\Omega  \ll \gamma $ the rectangular "pi pulse" defined by Eq.
(14) lasts much longer than the relaxation time $\gamma ^{ - 1} $.
Therefore during the most part of the pulse the initial state of
the system is already ignorable, and the difference between the
sequence of long "pi pulses" and the continuous pump field is
unimportant. The spectrum of radiation of the two-level system
under the influence of the continuous monochromatic field has been
calculated by Mollow \cite{M69}. For $\Omega  \ll \gamma $ and the
pump frequency that equals that of the transition, $\omega _0 $,
the power spectrum in our notation has the form
\begin{equation}\label{22}
P\left( \omega  \right) = \frac{{\Omega ^2 }}{{2\gamma ^2 }}\left[
{2\pi \delta \left( {\omega  - \omega _0 } \right) +
\frac{{2\Omega ^2 \gamma }}{{\left( {\Delta ^2  + \gamma ^2 }
\right)^2 }}} \right],
\end{equation}
where $\delta (z) $ is the Dirac's delta-function.  The first term
in the brackets describes the scattering of the pump radiation
with unchanged frequency.  Only the second, incoherent term can be
interpreted as a spectrum of the "spontaneous" emission by the
atom that is excited by a weak resonant field. The term
"spontaneous" in this case may be too stretched, since the atom in
the continuous monochromatic field can hardly be considered a
secluded one. However, if one accepts this interpretation of the
incoherent term, then it may be noted that for large $|\Delta |$
it follows the inverse quartic law, similar to our Eq. (20).
Pushing the interpretation further, we may say that for the
spectral distribution given by the incoherent term the Zeno time
and the quantum jump time happen to be equal: $\tau _Z = \tau _J =
\Gamma ^{ - 1} $.

In conclusion, we have demonstrated that the line shape of the
radiation, spontaneously emitted by an atom excited by a strong pi
pulse is Lorentzian only in the domain of frequency detunings that
do not exceed the Rabi frequency of the pulse, $\left| \Delta
\right| < \Omega $.  For larger values of $|\Delta |$  it falls
off much faster.  For the transition with the frequency $\omega _0
= 3.5 \cdot 10^{15} \,{\rm{s}}^{ - 1} $ and the momentum matrix
element $p_{12}  = 1.6 \cdot 10^{ - 20\,}
\,{\rm{g}}\,\,{\rm{cm}}\,\,{\rm{s}}^{ - 1} $  the spontaneous
emission rate is $\Gamma  = 1.3 \cdot 10^7 \,\textrm{s}^{ - 1} $.
The condition $\Omega  = 10\gamma $ corresponds to the intensity
of the rectangular pi pulse $I = 92\,\,{\rm{mW}}\,{\rm{cm}}^{ - 2}
$ and to its duration $\theta  = {\pi  \mathord{\left/ {\vphantom
{\pi  {\Omega  = 47\,{\rm{ns}}}}} \right.
\kern-\nulldelimiterspace} {\Omega  = 47\,{\rm{ns}}}}$.  In these
conditions for the Lorentzian line shape approximately 6\% of the
emitted photons must have frequency detunings $\left| \Delta
\right| > \Omega $.  The observation of shortage of these photons
seems to be accessible to the modern spectroscopic experiment.

The author is grateful to A.V. Borisov, G.A. Chizhov, L.V.
Keldysh, S.P. Kulik, A.E. Lobanov and E.A. Ostrovskaya for useful
discussions. The author acknowledges the support by the "Russian
Scientific Schools" program (grant no NSh-4464.2006.2).

\end{document}